# The Ties that Bind Networks:
# Weak Ties Facilitate the Emergence of Collective Memories

Ida Momennejad[a, b], Ajua Duker[a], Alin Coman[a]

[a]Department of Psychology. Princeton University, Princeton, NJ 08544

[b]Princeton Neuroscience Institute, Princeton University, Princeton, NJ 08544

## Abstract

From families to nations, what binds individuals in social groups is the degree to which they share beliefs, norms, and memories. While local clusters of communicating individuals can sustain shared memories and norms, communities characterized by isolated cliques are susceptible to information fragmentation and polarization dynamics. We employ experimental manipulations in lab-created communities to investigate how the temporal dynamics of conversational interactions can shape the formation of collective memories. We show that when individuals that bridge cliques (i.e., weak ties) communicate early on in a series of networked interactions, the community reaches higher mnemonic convergence compared to when individuals first interact within cliques (i.e., strong ties). This, we find, is due to the tradeoffs between information diversity and accumulated overlap over time. By using data calibrated models, we extend these findings to a larger and more complex network structure. Our approach offers a framework to analyze and design interventions in communication networks that optimize shared remembering and diminish the likelihood of information bubbles and polarization.

______________________________________________________________________

## Significance

Understanding the processes by which communities form collective memories has thus far eluded social scientists. By manipulating the sequence of conversational interactions in fully-mapped social networks we clarify the role that the network structure and temporal dynamics of conversational interactions plays in the formation of collective memories. As such, our approach offers a methodological framework to understand the formation of collective memories as a dynamical process. At the same time, it allows for exploring the emergence of information bubbles and attitude polarization in human communities and points to strategies for minimizing their negative impact on society.

## Introduction

Social interactions are crucial to communities that engage in coordinated behavior. These interactions constitute the main medium in which beliefs, memories, and norms, become shared across communities. They can facilitate the spread of information about healthy behaviors (1), change negative norms (2), and enable large-scale cooperation (3). On the other hand, when most individuals interact within homophilous social cliques (4, 5) they can give rise to information bubbles (6) and political polarization (7, 8) and can disrupt optimal collective behavior (9, 10). Understanding how the network structure of interactions affects such large-scale outcomes is crucial to optimizing the use of social networks for the collective good. Despite insightful new research into how communities form collective memories (11, 12), we know very little about the dynamical processes involved. Here, we show that the temporal structure of conversations in a social network increases large-scale convergence in the memories of its members.

We build on the extensive psychological research showing that once an experienced event is encoded, it's memory is malleable (13–15). It is subject to cognitive transformations, such as forgetting and distortion (16) and susceptible to social influences (17, 18). Due to this malleability, jointly remembering the past often leads to the synchronization of memories between interacting partners by strengthening discussed and suppressing non-discussed information (19). When these dyadic level influences are part of a larger network of social interactions, collective outcomes emerge (11, 17, 20, 21). The influence that one individual exerts over another can propagate through the network and impact the degree to which communities converge on a similar memory of an experienced event. Not all community members are, however, equally influential in their potential to affect the collective memory of the community. Individuals that connect between cliques (i.e., weak ties) – have been found to have a disproportionate influence on large scale outcomes (22–24). No research to date has experimentally explored the importance of weak ties on the convergence of individual memories across a network.

Crucially, social interactions within communities unfold over time. Depending on the sequential order of conversations, a 'weak tie' may never get the chance to impact the network, especially if it occurs after the community had already engaged in extensive interactions in isolated cliques (i.e., strong ties). Most previous investigations use static topological mappings to showcase the impact of weak ties (23, 25). In contrast with these approaches, we use a temporal network approach to study when 'weak tie' conversations should take place to maximally impact the convergence of memories across the network. To do so, we experimentally manipulate the temporal order of 'weak tie' and 'strong tie' conversations in laboratory-designed networks. We measure how this manipulation impacts collective level convergence of memories. Building on this empirical data, we propose a psychologically informed temporal network framework for predicting the mnemonic convergence of any given socio-temporal network and test this framework by using a real-world network (34).

We hypothesized that if participants that are connected through a weak tie discuss memories of a commonly shared event early on in the network's conversational cycle, they will facilitate widespread mnemonic convergence in the community. In contrast, early alignment between strong ties should lead to less mnemonic convergence across the community. To test this hypothesis, we keep the topological properties of conversational networks constant across experimental conditions, with all nodes having the same degree, closeness centrality, betweenness centrality, and eigenvector centrality. We only manipulate the temporal order of interactions within these networks so that the first round of conversations occurs either on the weak tie or on the strong tie (*see Fig.1*).

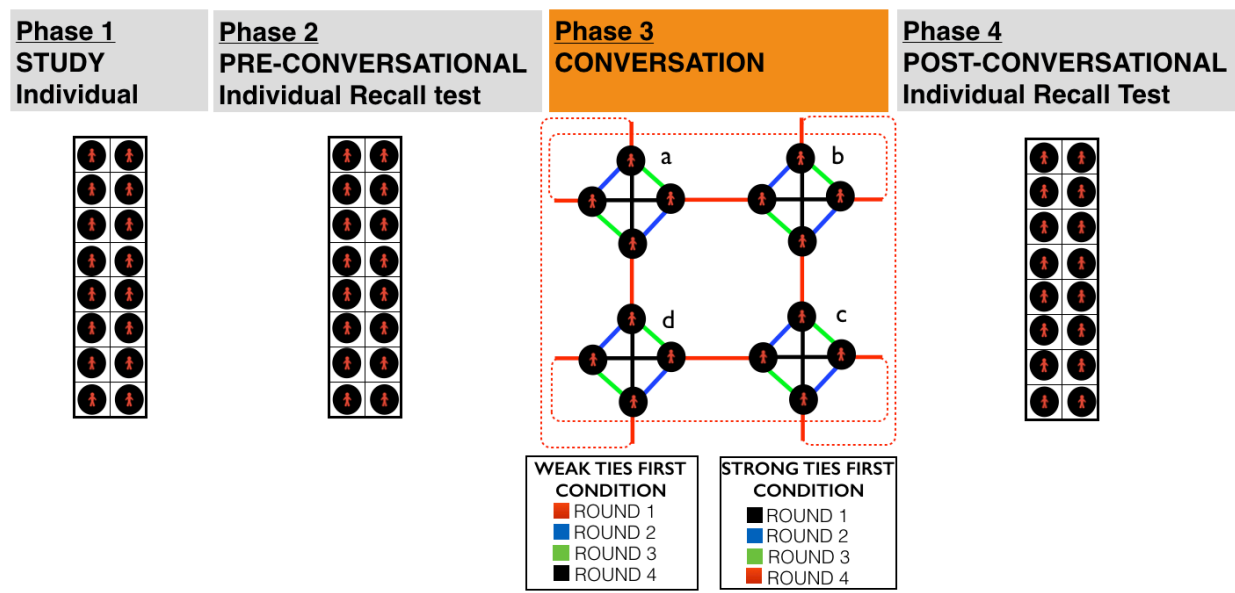

**Fig. 1. Phases of the experimental procedure**. The nodes represent human participants and the edges represent conversations between two individuals. Both experimental conditions shared the same static network topology. They only differed in the order of sequential dyadic interactions. In the Weak Ties First condition participants had their first conversation across cliques, whereas in the Strong Ties First condition they had their first conversation within cliques. The color of the links indicates the temporal sequence of the conversations; a, b, c, d indicate the predetermined cliques.

One hundred and ninety-two participants enrolled in the study through Princeton University's recruitment system. They were assigned to 16-member communities and completed the experimental procedure on lab computers. In the study phase (Phase 1), participants read a story that contained 30 critical items (16). Then, in the pre-conversational recall phase (Phase 2), they individually recalled the studied information. In the conversational recall phase (Phase 3), each participant in the 16-member community was paired for a series of four anonymous dyadic conversations (each with a different partner), during which they were instructed to jointly remember the studied materials. Conversations took the form of interactive exchanges in a chat-like, computer-mediated environment in which participants typed their recollections. Finally, in the post-conversational recall phase (Phase 4), they individually recalled the initially studied information (*Fig.1*).

In the conversational recall phase, each participant engaged in a sequence of four 150-s conversations. In the Weak Ties First condition (n = 96 participants; six 16-member networks), the conversational sequence began with interactions between individuals that belonged to different pre-determined cliques. In the Strong Ties First condition (n = 96 participants; six 16-member networks), the first interaction occurred between individuals who were part of the same pre-determined clique. The second and third interactions took place within cliques in both conditions, while the fourth conversation again differentiated between the Weak Ties First condition, in which participants now communicated within the clique, and the Strong Ties First condition, in which participants interacted between cliques (*see Fig. 1*).

Each person's memory was operationalized as a vector with 30 slots corresponding to the 30 critical items in the studied material, with 1 indicating than an item was recalled and 0 otherwise (*see Methods*). We computed mnemonic convergence scores for each 16-member community, separately for the pre-conversational and post-conversational individual recalls. To do so, we first calculated a mnemonic similarity score for each pair of participants by dividing the number of items the two participants remembered in common by the total number of non-overlapping items remembered by both participants (11). For each community, the mnemonic convergence score was calculated by averaging the mnemonic similarity scores across all pairs of participants in the network, separately for the pre-conversational and post-conversational recalls (*see Supplementary Fig. 1*). A mnemonic convergence score of 0 indicates that participants within a community remembered nothing in common, while a score of 1 indicates perfect overlap among all participants in the community.

## Results

*Dynamics of mnemonic convergence*. To explore whether weak ties impact the emergence of collective memories, we first compared mnemonic convergence scores in the two conditions. Consistent with our hypothesis, we observed a higher post-conversational mnemonic convergence in the Weak Ties First condition than in the Strong Ties First condition (*Fig. 2A*). This pattern, we argued, is due to the fact that weak ties lead to the increased similarity among individuals who belong to connected cliques. In order to investigate this claim we computed, separately for pre and post-conversational recall: (i) *within-clique similarity* by averaging the mnemonic similarity scores of participants who were part of the same clique (e.g., participants in Clique a in *Fig. 1*), (ii) *neighboring-clique similarity* by averaging the mnemonic similarity scores of non-interacting participants who belonged to adjacent cliques (e.g., participants from Clique a and participants from Clique b in *Fig. 1*), and (iii) *distant-clique similarity* by averaging the mnemonic similarity scores of participants who belonged to non-adjacent cliques (e.g., participants from Clique a and participants from Clique c in *Fig. 1*). For each of the three similarity types we subtracted pre from post-conversational scores for a measure of mnemonic similarity increase.

We found support for our prediction that weak ties affect network-wide mnemonic convergence by aligning the memories of individuals who are part of neighboring cliques. In the Weak Ties First condition, participants' memories were more similar to those from the neighboring clique participants than in the Strong Ties First condition. As predicted, in the Weak Ties First condition, memories propagated more efficiently across clusters (*Fig. 2B*). We predicted no difference between the two conditions in distant-cluster similarity, because the influence of one participant over another's memories can only propagate into the neighboring cluster and no further. The temporal structure of conversations makes it impossible for memory influences to manifest themselves between distant cliques, in either condition. Indeed, the mnemonic similarity increase was not statistically different between the Weak Ties First and the Strong Ties first conditions for distant cliques comparisons.

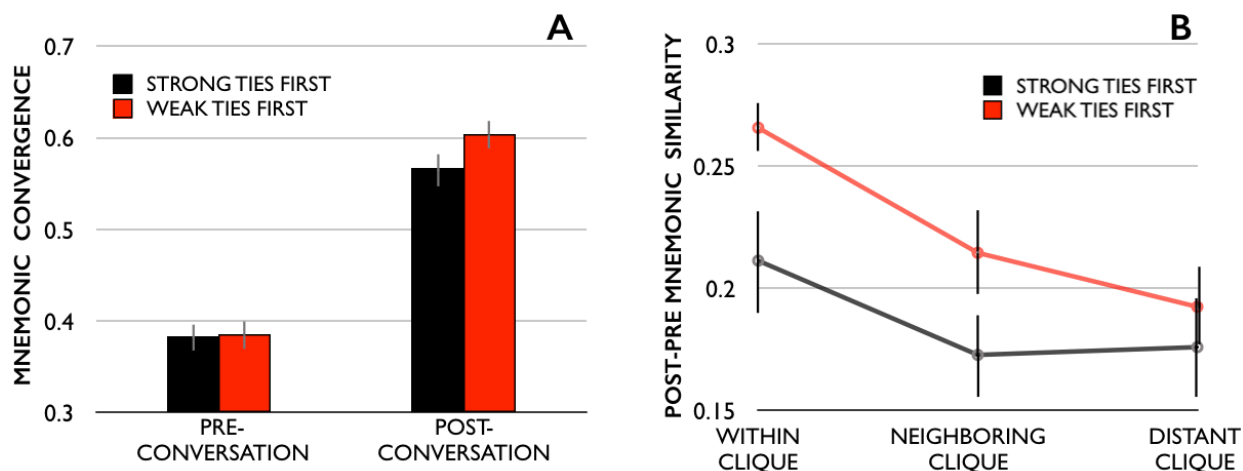

**Fig. 2. (A) Mnemonic convergence scores**. The increase in mnemonic convergence from pre-conversation to post-conversation was larger in the Weak Ties First condition (M=.22; SD=.03) than in the Strong Ties First condition (M=.18; SD=.03), P=.052, CI [.01,.07]. (**B) Clique-level Mnemonic Similarity scores**. The increase in mnemonic similarity from pre to the post-conversation was larger in the Weak Ties First condition than in the Strong Ties First condition for participants who were part of the same clique ($M_{wt}$=.27, $SD_{wt}$=.02 vs. $M_{st}$=.21; $SD_{st}$=.03; P=.006, CI [.02,.09]) and for participants in neighboring cliques ($M_{wt}$=.21, $SD_{wt}$=.04 vs. $M_{st}$=.17; $SD_{st}$=.03; P=.06, CI [.00,.09]), but not for those in distant cliques (P=.39). Error bars represent ±1 SEM.

*Differences in information diversity and accumulated overlap*. As to the alignment of memories among participants within-clique, two possibilities emerge. On the one hand, within-clique participants in the Strong Ties First condition might align their memories in the first round. By consolidating these memories in subsequent rounds, they might form locally convergent memories resistant to influence from neighboring clusters in the last round. On the other hand, this early consolidated alignment might be perturbed by the introduction of information from neighboring cliques in the final round. As each member of the clique aligns with a different "outsider" in the final conversational round, the clique may diverge from their locally formed collective memory in different directions. This account would predict lower within-clique convergence in the Strong Ties First compared to the Weak Ties First condition. Supporting the latter hypothesis, we found that participants in the Strong Ties First condition reached lower within-clique alignment than those in the Weak Ties First condition.

    We propose that higher within-clique convergence in the Weak Ties First condition is driven by the temporal dynamics of information diversity and accumulated overlapping recall (*Fig. 4*). High initial information diversity that gradually becomes shared among clique members in subsequent conversations might create the optimal conditions for within clique local convergence. In Round 1 of the Weak Ties First condition, each clique member had a conversation with an individual from another clique (see *Fig. 3A*). Thus, 8 individuals contributed to the pool of items collectively recalled in Round 1 by the clique. As a consequence, the pool of items that the clique remembered collectively should be characterized by high information diversity (i.e., number of items remembered in at least one, but no more than three conversations of

the participants in the clique) and low accumulated overlap (i.e., number of items remembered in all conversations of participants that form a clique). In contrast, in the Strong Ties First condition, the collective pool of information within a clique should be characterized by low information diversity and high accumulated overlap, because their Round 1 interactions are within the clique (4 sources of memory contributed to recall).

To test this hypothesis, we computed two indices that characterize the information that is available to each clique: an information overlap index and an information diversity index. For each participant within a clique, and for each conversational round, we constructed a 30-item vector to capture the information produced in each participant's conversation. Because each cluster involves 4 participants, there were four vectors that went into the computation, corresponding to each individual's vector. If an item (among the 30 initially studied) was present in all four vectors, it was designated as an overlap item. The total number of overlap items constituted the information overlap index for the clique. If an item was present in at least 1 vector, but in no more than 3 vectors, it was designated as a diversity item (*Fig. 3A*). The total number of diversity items constituted the information diversity index for the clique. These indices were computed for each clique and then averaged across the four cliques in a network.

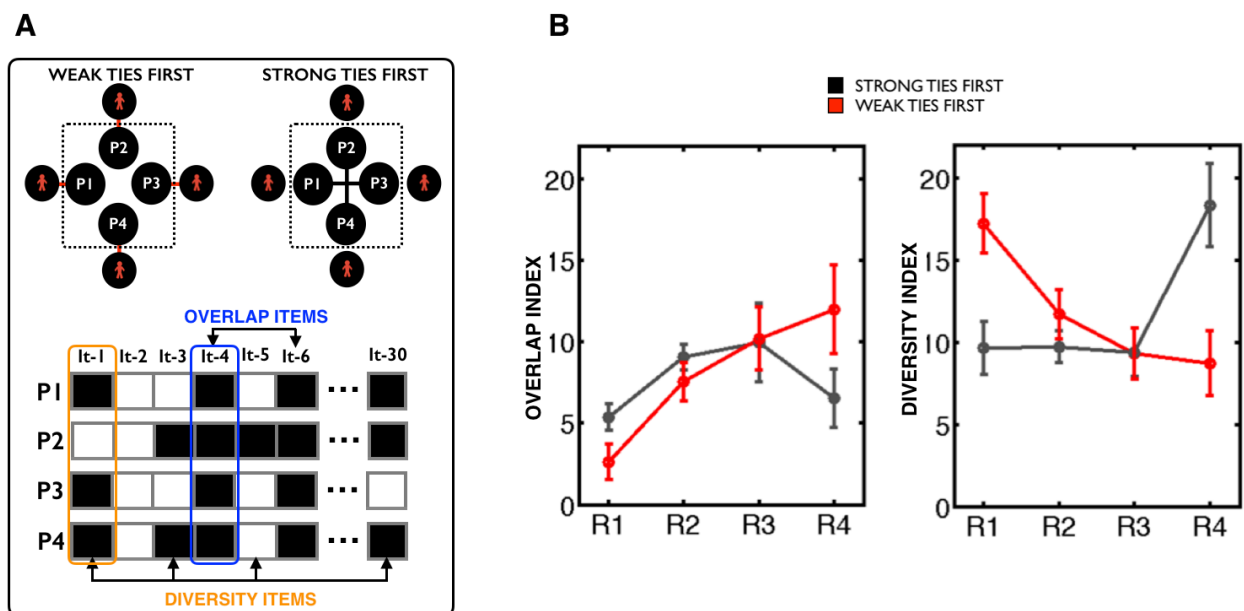

**Figure 3. Recall overlap and diversity indices**. (A) For each clique (designated with the dotted line), we compared vectors corresponding to information that came up in conversations of each of its 4 members (P1-P4). Some of these conversations were within-clique, and some were between a clique member and an "outsider" (red figures). For each round, a cluster's overlap index indicated the number of items (out of the 30 studied items) that came up in conversations of all members (highlighted with the blue box), while the diversity index indicated the number of items that were mentioned in at least one but no more than three conversations (highlighted with the orange box). (B) The dynamics of the overlap and diversity indices, by round, separate for the Weak Ties First (red) and Strong Ties First (black) conditions. The interaction between Time and Condition is significant for both the overlap (p=.003) and the diversity (p=.001) indices, with post-hoc analyses indicating significant differences between the Weak and Strong Ties First conditions in both Round 1 (overlap index: p=.002; diversity index: p<.001) and Round 4 (both indices p<.001). Error bars represent ±1 SEM.

Consistent with our hypothesis, we found that in the Weak Ties First condition, the information was initially (Round 1) less overlapping and more diverse than in the Strong Ties First condition. As participants engaged in subsequent conversations, this dynamic reversed, such that information brought up in the last round was more diverse and had a lower overlap score in the Strong Ties First condition than in the Weak Ties First condition. This dynamic, we argue, resulted in higher within-clique alignment in the Weak Ties First condition, but had the opposite effect in the Strong Ties First condition, where the information that was introduced in Round 4 had no time to be integrated and discussed within the clique (*Fig. 3B*).

*Modeling the temporal dynamics of mnemonic convergence*. Experimenting with lab-created communities allowed us to exercise control over the temporal sequencing of conversational interactions, a critical feature needed to test our hypothesis. We next sought to develop a graph theoretic approach that would be calibrated on our empirical data and could be used to predict mnemonic convergence within any network, given knowledge of the temporal order of interactions.

We first derived general principles of memory influence by focusing on two memory effects that are well-established in the extensive psychological literature on memory (27, 28): the recency effect and the practice effect (29). We reasoned that a person's post-conversational memory would be influenced by (a) a recency parameter ($\lambda$), determining how influential the more recent conversations are for the participant's memory, and (b) a practice effect reflected in parameter ($\gamma$), determining the likelihood that a participant would propagate items from a given conversational round into subsequent rounds. Essentially, this practice effect represents the fact that items mentioned in earlier conversations are practiced at a higher rate than items mentioned in later conversations (29). We capture this principle as an exponential decay in the practice rate (probability of propagation) as rounds go by.

To calculate the influence any node could have on any other node in the network we first calculate temporal path lengths of propagation (30–32). Multiplying adjacency matrices of each round allows the model to capture all mnemonic paths between 2 nodes. For example, let us consider the adjacency matrices displayed in *Fig. 4A*. In the first round of conversations (captured in adjacency matrix A1) person 1 influences person 4, and in the second round (A2) person 4 influences person 3. Matrix multiplication (A1XA2) reveals that, even though persons 1 and 3 have never been directly in contact, person 1 has influenced person 3, via person 4, by the end of the second round. As such, by the end of round 2, there is a temporal distance of 2 from person 1 to person 3, but an infinite temporal distance from person 3 to person 1, and hence a mnemonic influence of zero (because person 3 has not influenced person 1). We then compute a 'mnemonic reachability' score between every two nodes by combining principles of matrix operation in temporal networks and applying the recency and practice parameters estimated from our empirical data to the temporal network as follows (Equation 1):

$$C = \sum_{t=1}^{r}(1-\lambda)^{r-t} A \ + \sum_{t=1}^{r-1} \sum_{k=t+1}^{r} \prod_{i=t}^{k} \gamma^k A_i$$

Here $A_x$ refers to the adjacency matrix of conversations in round x, and *r* refers to the maximum number of rounds (in this experiment *r=4*, and $A_4$ indicates the adjacency matrix of conversations at round 4). The parameter $\gamma$ captures the conversational practice rate, corresponding to the rate of propagation of a person's conversational recall from one round to subsequent rounds. It has a value between 0 and 1 and exponentially decays on each round *t* according to $\gamma^t$. The parameter $\lambda$ is the recency rate, with a value between 0 and 1. The larger $\lambda$ is, the more the post-conversational recall will resemble the most recent conversation rounds (i.e., strong recency effect).

To calibrate the model with the empirical data we first applied *Equation 1* to the adjacency matrices of each conversation round in the 16-member networks by using all the possible range of values between 0 and 1 (in 0.1 increments) for our two parameters (i.e., $\lambda$ and $\gamma$). This produced different 16 X 16 matrices of reachability for all the different ranges ($M_{Model}$). We computed the correlation between the 16 X 16 matrix corresponding to the mnemonic similarity scores observed in the human data ($M_{Human}$) and the model matrices generated using various values for $\lambda$ and $\gamma$. Using this method, we identified the values of the parameters that maximized this correlation (i.e., values for $\lambda$ =0 to 0.2, values for practice rate, $\gamma \cong .5$), thus calibrating our model with the empirically obtained data (see *Supplementary Fig. 3*).

*Validating the model with real-world social networks.* To validate our model, we selected a real-world social network that captures the face-to-face interactions of people during an exhibition in 2009 at the Science Gallery in Dublin on infectious diseases (i.e., Infection exhibition network) (33). Nodes represent exhibition visitors; edges represent face-to-face contacts that were active for at least 20 seconds, as measured with an electronic monitoring device (*see Fig. 4C*). The network contains the data from the day with the most interactions (5330 interactions) among 410 nodes. Next we applied *Equation 1* to the Infection Exhibition network. We used the Girvan-Newman edge betweenness algorithm (4) to sort the edges into 4 quartiles. Using a descending order of betweenness scores, we derived 4 rounds of 1382 conversations each, corresponding to the Weak Ties First condition.

We then swapped the first and the last quartile of conversations based on betweenness score to arrive at a Strong Ties First temporal structure. We employed spectral clustering (34, 35), using a Fiedler vector (the second smallest eigenvector of the graph Laplacian), to identify within-cluster, neighboring-cluster, and distant-cluster relations. We applied *Equation 1* to the adjacency matrices of the Infection Exhibition network, and plotted the predicted scores for the two conditions (*Fig. 4C*). We also computed convergence for a random edge order, allowing for weak and strong ties to happen in every one of the four conversational rounds. We observed the that this random edge-order predicted a higher level of convergence compared to either descending tie strength (our approximation of Weak Ties First condition) or ascending tie strength (our approximation of the Strong Ties First condition).

    **A**        **Model**

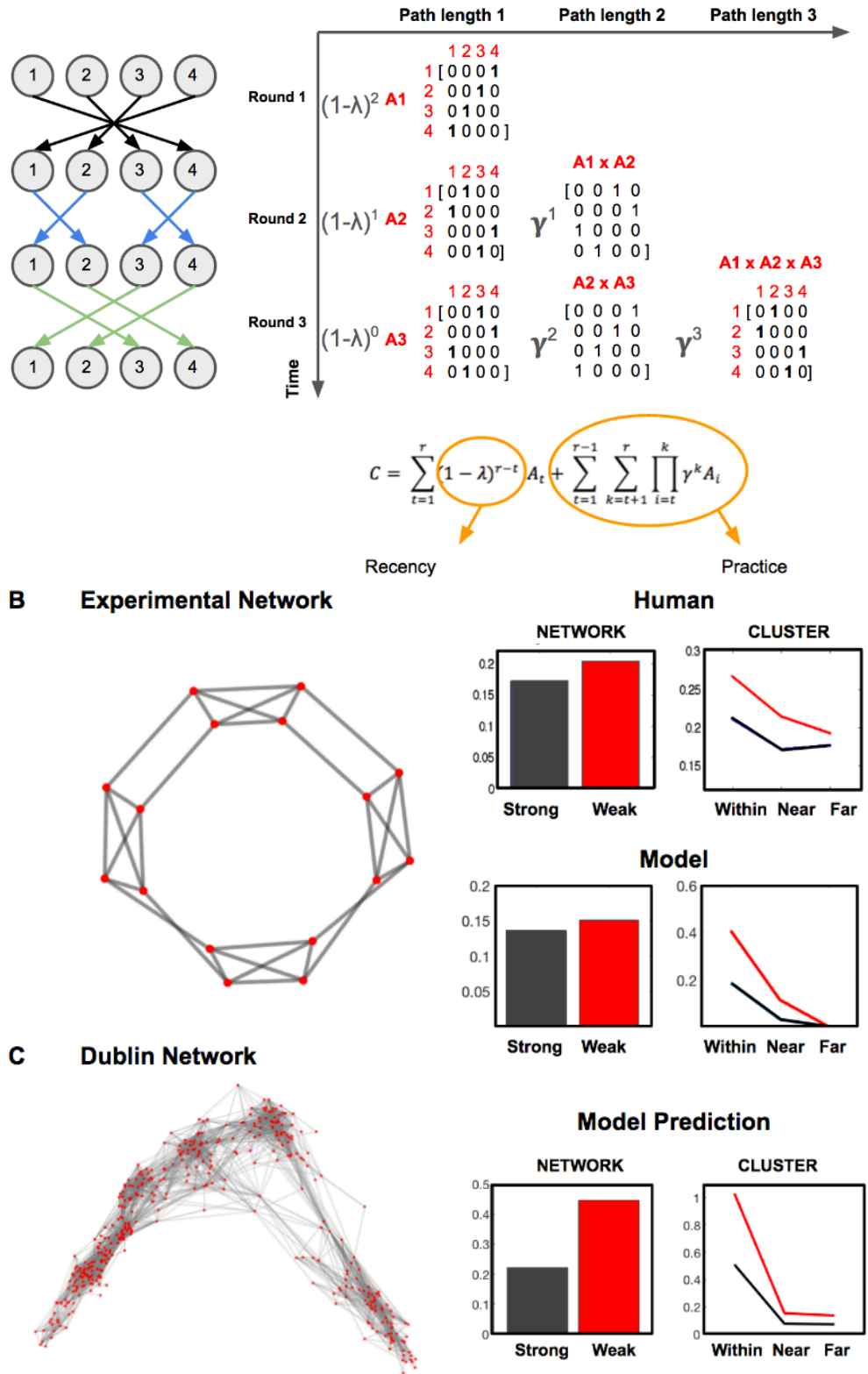

**Fig. 4. Human performance and model prediction.** (A) A representation of the computation of the reachability scores between 2 nodes over time; the recency effect parameter and the practice effect parameter are applied as a function of temporal sequencing of each round. In the figure, numbered nodes represent participants and arrows represent conversations that take place in different rounds. For simplicity, we only illustrate 3 rounds of conversations (B) We applied Equation 1 to the adjacency matrices of each conversational round and a range of propagation decay parameters. As such, for each node a score of mnemonic reachability to each other nodes is computed. The score takes into account the temporal network of interactions and integrates the recency and practice effects. We then averaged the reachability scores across the nodes that comprise a network, separated by condition (bar graph). We also computed specific scores by averaging these reachability scores depending on whether the nodes belonged to the same clique, neighboring cliques, or distant cliques. We compared these scores with our behavioral results, i.e. the person-to-person matrices of mnemonic similarity for the experimental conditions. These comparisons suggest that it is possible to predict mnemonic convergence and similarity scores by using these reachability scores. (C) To test whether the model predicts similar Weak Tie First effects in a real-world network as in our lab experiments, we applied equation 1 to an existing network of human interactions (with 410 nodes and 5330 edges). The model made similar predictions to the findings of the 16-person network.

Taken together, these findings suggest that allowing weak ties to communicate early on, either all at once or intermixed with strong ties, leads to more convergent collective memories compared to a condition when the strongest ties take place first. Consistent with the human data, the predicted within-cluster convergence was higher in the weak ties first condition than in the strong ties first condition, thus establishing the generality of this phenomenon.

## Discussion

Previous research has shown that network structure impacts the emergence of collective memories in communities of individuals (11). The approach developed here provides two principal advances in our understanding of these emergent processes. First, we offer a framework for quantifying the temporal dynamics involved in the formation of collective memories. We show that the same network topology, depending on the temporal sequencing of conversations, gives rise to different levels of convergence. Second, we demonstrate that optimizing the temporal dynamics of cross-cluster interactions, via weak ties, can enhance mnemonic convergence. We have shown experimentally, for the first time, that individuals that bridge network clusters have a strong influence on the emerging collective memory in lab-created communities.

In the current study we systematically manipulated the sequence of conversational interactions. The principles extracted from this experimental investigation could be easily applied to real-world social networks. On the one hand, one can use these principles to predict the level of convergence real-world communities might reach given observed topological and temporal features of their networked interactions. On the other hand, in situations when control could be exercised over the sequencing of social interactions (e.g., classrooms, organizations, etc.), one could design these interactions and optimize for desired outcomes, such as mnemonic convergence, information homogeneity and diversity, and recall accuracy.

We offer a practical application of temporal networks (36) in a model that combines graph theory with well-established memory effects to predict mnemonic convergence. We have selected values for practice and recency parameters that maximize the correlation between post-conversational convergence matrices of the model and human data. As shown, our model's predictions generalize to larger and more realistic socio-temporal networks (e.g., Dublin network). However, further research in this direction is certainly warranted. The literature on collective memory would benefit from systematic investigations of the impact of network size and structure on the dynamical transformation of memories (37). We are just beginning to understand how the malleability of human memory impacts the formation of collective memories. There are a multitude of factors that could interact with the temporal and topological features of social networks to give rise to collective memories. People's attitudes (38), the perceived similarity among interactants (39), and the medium in which interactions take place (40, 41) likely play a role in the synchronization dynamics that characterize social groups. In future work, we aim to use experiments with larger and more realistic social networks to investigate the impact of these factors under ecologically valid circumstances.

Ultimately, understanding collective cognition will clarify social phenomena that underlie significant challenges of our times: information bubbles and extreme attitude polarization. Highly clustered communities, in which individuals reinforce their memories in repeated within-cluster interactions lead to fragmented collective memories. This, in turn, could produce polarization, largely because the different factions rarely exchange information to reach common ground (42). Our proposed approach can help strategically optimize information ecosystems for maximal knowledge acquisition and efficient community organization. These predictions would provide the grounding for interventions to avoid information bubbles and reduce the spread of misinformation in communities.

## Materials and Methods

*Participants.* A total of 192 students (123 female, mean age 21.83 years, with a 3.97 SD) affiliated with Princeton University took part in the study voluntarily for either research credit or compensation. They were grouped into twelve 16-member communities and went through the study in a Princeton computer lab, which contains visually partitioned computers. The participants interacted anonymously through the software SoPHIE (Software Platform for Human Interaction Experiments); they were from a wide range of fields of study which made it unlikely that any subject would know more than one other person in the room. We asked subjects for their sex and major field of study. In total, 12 sessions were conducted, 6 in the Weak Ties first condition and 6 in the Strong Ties first condition. All subjects gave informed consent for the protocol, which was approved by Princeton University's Institutional Review Board.

*Stimuli.* Using the Qualtrics survey paradigm, we presented participants with a story taken from Anderson & Pritchard (1978) (16). The 30-items story contains information about two boys who skip school and visit one of the boy's house. Even though the story was initially designed (16) to contain items relevant for two cognitive schemas (i.e., real-estate mindset vs. burglar mindset), our manipulation did not involve the activation of these schemas.

*Design and Procedure.* Participants signed up for the study through Princeton University's online recruitment system. Each session was conducted with 16 participants who went through the experimental procedure together. Participants within each network were physically present in the same room and carried out the study on the designated computer terminals throughout. In the study phase, participants initially studied the story for a fixed amount of time. They were told that their memory will be tested in a later phase. Then, in a pre-conversational recall phase, participants were asked to individually remember as much as they could about the initially presented information. After this phase, participants took part in a sequence of conversations for which they were instructed to jointly remember the initially studied materials (conversational recall). These computer-mediated chat conversations took place in dyads, such that each participant in a 16-member community engaged in a sequence of four interactions. The conversations were characterized by turn-taking, with virtually all conversational recall instances involving collaboration between the interacting partners. A qualitative analysis of the conversations revealed that all of the participants stayed on task throughout the duration of the study and engaged in collaboratively remembering the initially studied materials, as instructed.

     We manipulated the network structure of conversational interactions as illustrated in *Fig. 1*. A software platform was specifically designed for this project to allow for fluent computer-mediated interactions among participants (i.e., SoPHIE). We kept the number of participants and the number of conversations in which each participant was engaged constant across the two conditions. The only difference between the two conditions was the sequencing of conversational recall sessions. In the Weak Ties first condition (6 communities) the first conversation occurred between participants who were part of different clusters, while in the Strong Ties first condition (6 communities), the first conversation took place between participants who were part of the same cluster. Each conversation lasted for 150 seconds, which preliminary studies

showed that it provided ample time for information to be exchanged. A final recall test followed the conversational phase (post-conversational recall). Coding of all of the recall protocols was performed by a research assistant who was blinded to the study's hypotheses and involved a binary system in which an item was labeled as either remembered or not remembered. The coding scheme followed the designation employed in the original study by Anderson & Pritchard (1978). Ten percent of the data were double-coded for reliability (Cohen $\kappa=.84$). Three- to 5-min distracter tasks, in which participants completed unrelated questionnaires, were inserted between any two phases described above.

**SUPPLEMENTARY FIGURES**

| MEASURE/DEFINITION | FIGURE | FORMULA |
|---|---|---|
| **MNEMONIC SIMILARITY** Similarity between the memories of any two participants, computed both pre and post-conversation. | 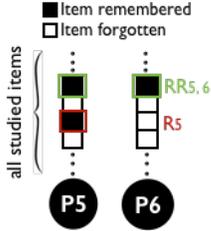 | $MS_{i,j} = \dfrac{R_i \wedge R_j}{R_i \vee R_j}$ $R_i \wedge R_j$ - number of items participants i and j remembered (black squares) in common $R_i \vee R_j$ - the non-overlapping number of items participants i and j remembered in total |
| **MNEMONIC CONVERGENCE** Average of mnemonic similarity scores across a network, computed both pre and post-conversation. | 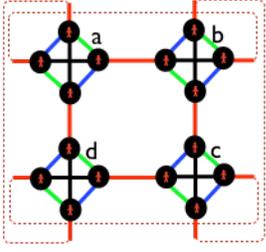 | $MC_{i:j} = \dfrac{\sum_{j}^{i} MS_{i:j}}{N_{i:j}}$ $MS_{i:j}$ - mnemonic similarity for all i-j pairs in a network $N_{i:j}$ - number of i-j pairs |

**Supplementary Figure 1**| A summary of definitions, figures, and equations for the dependent variables are presented.

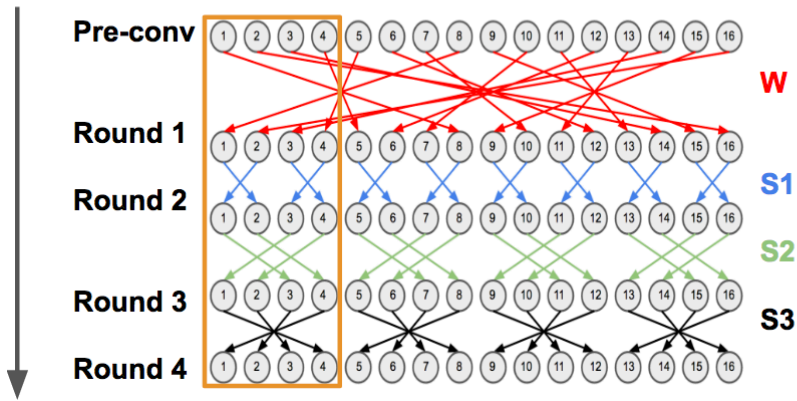

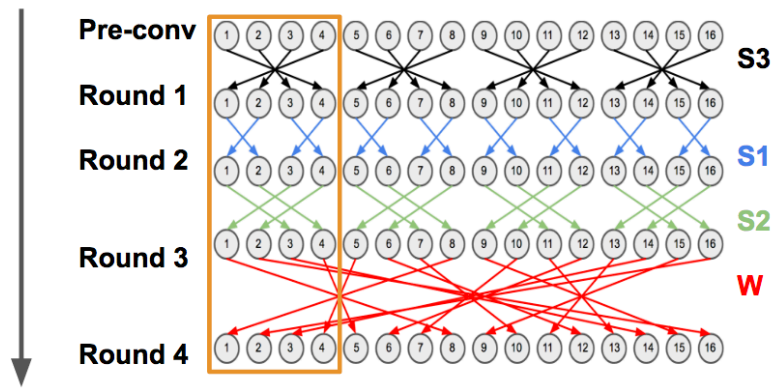

**Supplementary Figure 2 | Schematic of matrix operations with $\lambda = 0$.** This figure expands Equation 1 with one of the optimal $\lambda$ parameters in order to clarify the practice effects. Notably, we compared the results of Equation 1, graphically expanded for Weak Ties First and Strong Ties First temporal networks, to Equations proposed for assessing the diffusion of novel information in temporal networks. We found that the general temporal network equations do not address the emergence of collective memories. The reason is that these studies do not take into account the significance of memory effects such as forgetting (recency) and practice effects that we apply here.

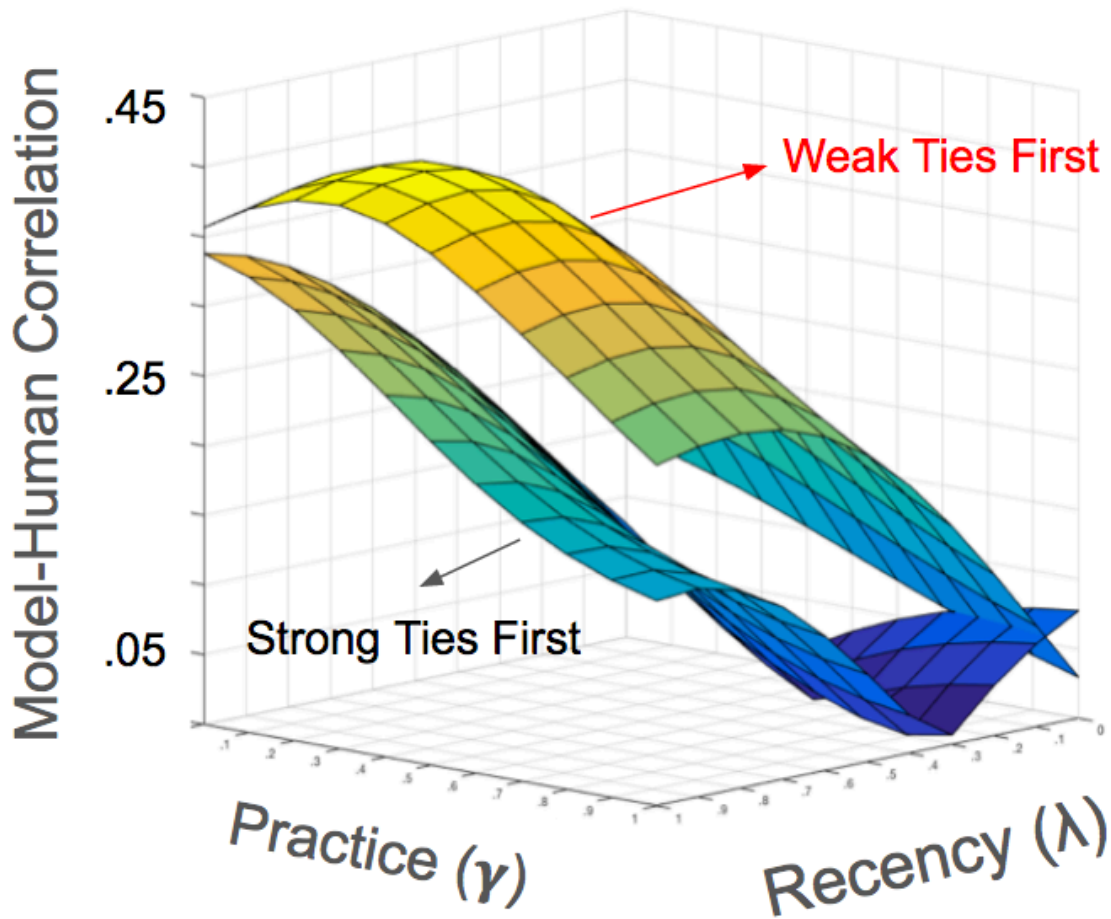

**Supplementary Figure 3 | Parameter selection.** In order to identify the optimal $\gamma$ and $\lambda$ parameters, we computed the correlation between matrices with the model's predicted similarity between nodes and matrices of post-pre memory convergence in humans. The human matrices were generated based on 12 networks each with 16 members, where conversations in 6 networks were organized in a weak ties-first fashion and the other 6 networks had conversations in the strong ties-first order. We observed that a recency or $\lambda$ value of 1 and a practice or propagation rate of $\gamma = .5$ resulted into the most qualitative fit to the data. We thereafter used $\gamma = .5$ and $\lambda = 1$ for all model predictions, including those applied to compute model predictions for the Dublin network (n=410).

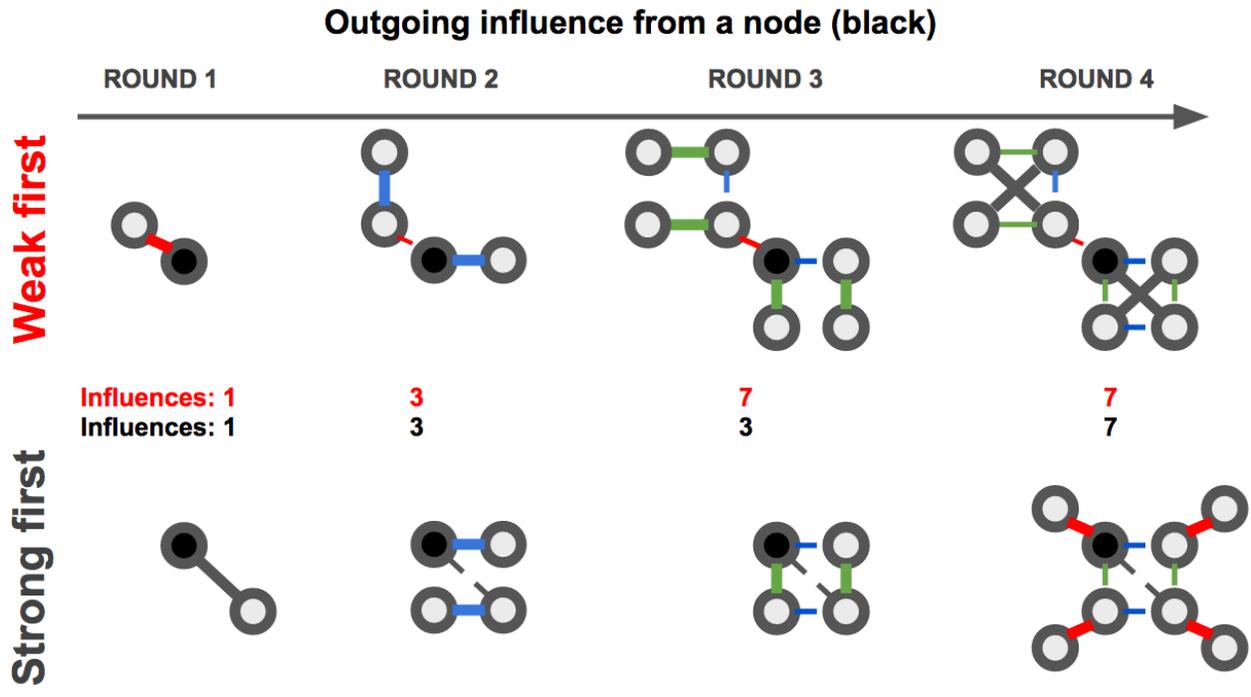
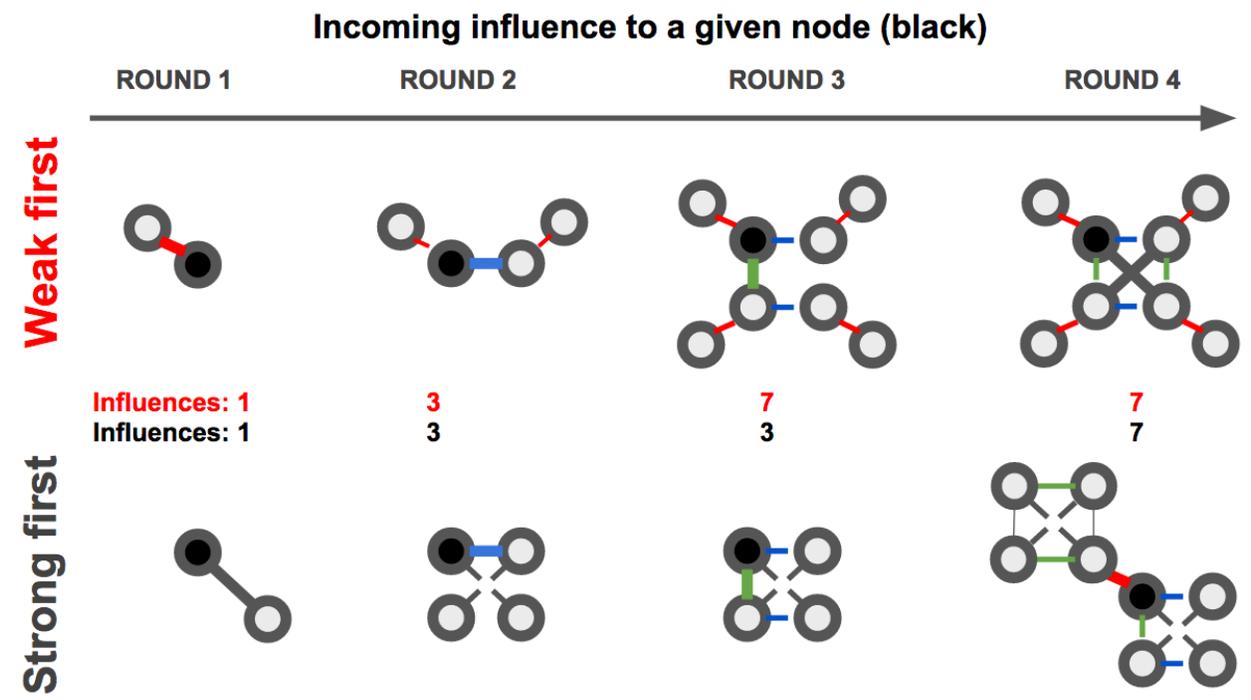

**Supplementary Figure 4 | The evolution of influence of and to a given node (shown in black).**